 \documentclass[preprint2]{aastex}

\newcommand{\bm}[1]{\mbox{\boldmath $#1$}}
\newcommand{\kms}{km s$^{-1}$}

\shorttitle{Evolution of Star Clusters near the Galactic Center}
\shortauthors{Fujii et al.}

\begin{document}

\title{Evolution of Star Clusters near the Galactic Center: Fully
Self-consistent $N$-body Simulations}


\author{M. Fujii\altaffilmark{1,3}, M. Iwasawa\altaffilmark{2,3},
Y. Funato\altaffilmark{2}, and J. Makino\altaffilmark{3}}

\altaffiltext{1}{Department of Astronomy, Graduate School of Science, The
University of Tokyo, 7-3-1 Hongo, Bunkyo, Tokyo 113-0033;
fujii@cfca.jp}

\altaffiltext{2}{Department of General System Studies, College of Arts and
Sciences, The University of Tokyo, 3-8-1 Komaba, Meguro, Tokyo 153-8902; 
iwasawa@margaux.astron.s.u-tokyo.ac.jp, funato@artcompsci.org}

\altaffiltext{3}{Division of Theoretical Astronomy, National
Astronomical Observatory of Japan, 2-21-1 Osawa, Mitaka, Tokyo,
181-8588; makino@cfca.jp}

\begin{abstract}
We have performed fully self-consistent $N$-body simulations of star
clusters near the Galactic center (GC).  Such simulations have not
been performed because it is difficult to perform fast and
accurate simulations of such systems using conventional methods.
We used the Bridge code, which integrates the parent galaxy using the
tree algorithm and the star cluster using the fourth-order Hermite
scheme with individual timestep.  The interaction between the parent
galaxy and the star cluster is calculate with the tree algorithm. 
Therefore, the Bridge code can handle both the orbital and internal
evolutions of star clusters correctly at the same time.  We
investigated the evolution of star clusters using the Bridge code and
compared the results with previous studies.  We found that 1) the
inspiral timescale of the star clusters is shorter than that obtained
with "traditional" simulations, in which the orbital evolution of star
clusters is calculated analytically using the dynamical friction
formula and 2) the core collapse of the star cluster increases the
core density and help the cluster survive.  The initial conditions of
star clusters is not so severe as previously suggested.
\end{abstract}


\section{Introduction}

A few dozens of very young and massive stars have been found in the
central parsec of the Galaxy \citep{K95,Pa01,Pa06}.  These stars are a
few million years old \citep{Pa01,Ghez03} and lie on a disk
\citep{Lu06} or two disks \citep{Pa06}.  The disks rotate around the
central black hole (BH) and are at large angles with each other. One
disk rotates clockwise in projection, the other counterclockwise
\citep{Pa06}. These disks are coeval to within 1 Myr.
The orbit of the stars on the clockwise rotating disk are circular
\citep{Pa06} or eccentric have lower-limit eccentricities of
$0.0-0.8$ \citep{Lu06}, while those on the counterclockwise disk have
high eccentricities at around 0.8 \citep{Pa06}.
In the central
parsec, in situ formation of these stars seems difficult because of
the strong tidal field of the central BH.  To overcome this
difficulty, two possibilities have been suggested: (1) in situ star
formation in a massive accretion disk, or (2) inspiraling young star
clusters.

The accretion disk scenario was proposed by \citet{LB03}.  A dense
gaseous disk is formed from a molecular cloud which somehow fell to
the neighborhood of the central BH. If the disk is sufficiently
massive it can become  gravitationally
unstable, resulting in  fragmentation and formation of stars.
However, this scenario is problematic. Observations have
shown that two disks are at large angles with respect to each other
and these stars on the disks formed almost at the same time. In this
scenario, two disks must have existed simultaneously within 1
Myr. Moreover, it is difficult to make stars with eccentric orbit from
accretion disks \citep{Nay07}.

The star cluster inspiral scenario was proposed by \citet{Gerhard01}.  A
star cluster was formed at a distance of tens of parsecs from the GC and
spiraled into the GC due to the dynamical friction.
This scenario is supported by the observational fact that two young
dense star clusters, the Arches and Quintuplet clusters are observed at
the distances of $\sim$ 30 pc from the GC\citep{Fi04}.  Although this
scenario can explain two stellar disks without difficulty, numerical
simulations have shown that it would take too long time for the star
cluster to inspiral to the central parsec unless it was very massive or its
initial position is very near from the GC (Portegies
Zwart et al. 2003, hereafter PZ03; G\"{u}rkan \& Rasio 2005).

In these works (PZ03; G\"{u}rkan \& Rasio 2005), the orbit of the
cluster within its parent galaxy was calculated using the dynamical
friction formula \citep{Ch43}. With this approach, the inspiral
timescale might have been overestimated.  In \citet{Fj06}, we
performed fully self-consistent $N$-body simulations of a satellite
galaxy within its parent galaxy and found that the orbital decay of
the satellite is much faster than those calculated analytically from
the dynamical friction formula.  This difference was caused by
particles escaped from the satellite. One mechanism is that
the direct gravitational forces from escaped particles worked as
effective drag force to the satellite. The second mechanism is that
escaped particles remain close to the body of the satellite and
enhance the dynamical friction. These effect should also work in the
case of star clusters. Therefore, a fully self-consistent $N$-body
simulation is necessary to obtain correct results for the orbital
evolution of star clusters.

Kim \& Morris (2003; hereafter KM03) performed
self-consistent $N$-body simulations of star clusters near the GC. 
Their results showed that if the initial central density of a star 
cluster was initially very high ($\sim 10^8 M_{\sun}$pc$^{-3}$), the 
cluster can deliver stars to the central parsec of the Galaxy.
In these simulations, however, 
the internal evolution of the star clusters was neglected. The
stars in their model of star clusters have an equal mass and a large
softening length of 0.025pc. Such star clusters experience neither mass
segregation nor core collapse.  However, if the core collapse occurs, 
the core density of a star cluster increases.  The initial high density
of the core will not be necessary.  Thus, it is also important
to solve the internal evolution correctly.

Such a fully self-consistent $N$-body simulation has been impossible
with conventional numerical methods.  While star clusters need a very
accurate scheme such as the combination of fourth-order Hermite scheme
and direct force calculation, galaxies contain too many particles to
use the direct force calculation.  To solve this problem, we have
developed a new tree-direct hybrid scheme, the ``Bridge'' scheme
\citep{Fj07}. The Bridge scheme enables us to perform fully
self-consistent $N$-body simulations of star clusters within their
parent galaxies in a realistic time (less than 2 days with a single
GRAPE-6 board).

We performed fully self-consistent $N$-body simulations of evolution of
star clusters within their parent galaxies using the Bridge scheme.  We
also performed the ``traditional'' $N$-body simulations, in which the
orbital decay of the star cluster is calculated using the dynamical
friction formula for comparison. We found that the inspiral timescale of
the star cluster is shorter than that obtained in previous studies. In
addition, if the initial orbit of the star cluster is eccentric, the
timescale of inspiral is much shorter than that for a cluster in the
circular orbit of the same apocenter distance.

The eccentricities of the stars
escaped from the star cluster distribute around the eccentricity of the
star cluster. Thus, if a star cluster is initially in an eccentric
orbit, it naturally explains the rather high eccentricities of  the
stars in the observed "disks". Also, because of the mass segregation
effect, very massive stars (more than $10M_{\odot}$) remained in the
cluster and were brought very close to GC. Thus, we conclude that the
timescale problem with the cluster inspiral scenario was partly because of
the wrong treatment of the dynamical friction and partly because the
limited assumption of the circular orbit, and it is not difficult to
make the central stellar disks from inspiraled star clusters.

We describe the simulation method and initial conditions in section
2. In section 3 we show the results of simulations. 
Section 4 is for summary and discussions.

\section{Numerical Simulation}

\subsection{Models}

We adopted a King model with $W_0=3$ for the model of a star cluster.
Its core radius, $r_{\rm c}$, half-mass radius, $r_{\rm h}$, and tidal
radius, $r_{\rm t}$, are 0.087 pc, 0.13 pc, and 0.47 pc, respectively.
It consists of 65536 stars and we assigned each star a mass randomly
drawn from a \citet{Sal55} initial mass function between 0.3 and 100
$M_{\sun}$, irrespective of position.  The total mass of the star
cluster, $M_{\rm SC}$, is $7.9\times 10^{4}{M_{\sun}}$.  This model
imitates the Arches cluster \citep{Nagata95}, whose mass and velocity
dispersion are $7\times 10^4{M_{\sun}}$ within 0.23 pc and 22 \kms,
respectively \citep{Fi02}. The Arches cluster is located at $\sim 30$
pc from the GC. We, however, placed our cluster much closer to the
GC, to compare our result with those of  PZ03.  In table
\ref{tb:models}, we summarize the model parameters.

\begin{figure}
\epsscale{0.8}
\plotone{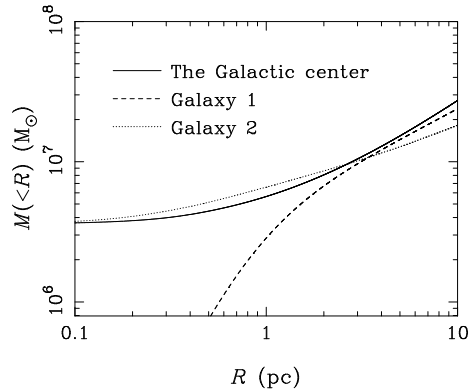}
\caption{Enclosed mass of our model galaxy and the GC \citep{Genzel03}.}
\label{fig:model}
\end{figure}

We used two galaxy models as a model of the central region of 
the Galaxy; one includes the central SMBH 
(galaxy 1) and the other does not (galaxy 2).
For the galaxy model 1, we adopted a King model with non-dimensional 
central potential of $W_0 = 10$.  We scaled the
density and velocity dispersion of our model at 5 pc from the GC to
the those of the Galaxy.  The density at 5 pc is $6.8\times 10^3
M_{\sun}{\rm pc}^{-3}$ and the one-dimensional velocity dispersion is
64 km s$^{-1}$ in our scales, while the observed density at 5 pc is
$\rho (5{\rm pc}) = 6.9 \times 10^3 M_{\sun}{\rm pc}^{-3}$
\citep{Genzel03} and the observed velocity dispersion is 54 \kms at
$\sim$ 4 pc \citep{Genzel00}. Figure \ref{fig:model}
shows the enclosed masses of our model galaxy and the result of
\citet{Genzel03}.  In this model, the total mass of the galaxy, $M_{\rm
G}$, is $8\times 10^7 {\rm M_{\sun}}$, and the core radius, $r_{\rm c,
G}$, and the half-mass radius, $r_{\rm h, G}$, are 0.66 pc and 21 pc,
respectively.  We used $2 \times 10^6$ particles to model the
Galaxy. The mass of a particle of the galaxy is 40 $M_{\sun}$. Since our
galaxy model has the finite core size of 0.66pc, the orbital evolution
of the star cluster should be reasonably accurate as far as its distance
from the GC is more than 1pc.  

The galaxy model 2 is a more realistic model of than model 1.  
It includes a central super-massive black hole (SMBH).  This model is 
based on King model.  We put a SMBH at the center of the galaxy with 
King model $W_0 = 10$ and integrated it for around two crossing times.  
The central region of the galaxy evolved and its density profile became 
cuspy.  This model roughly represents the Galactic center between
0.1 and 5 pc.  The enclosed mass is shown in figure \ref{fig:model}.
The total mass and the particle mass of the galaxy are 
$2.9\times 10^7 {\rm M_{\sun}}$ and 14 $M_{\sun}$, respectively. 

The initial position and velocity of the star cluster are shown in table
\ref{tb:initial}. We calculated two orbits; one is circular (model C) and 
the other is eccentric (model E) for the galaxy model 1, and two 
eccentric orbits (model B1, B2) for the galaxy model 2. 
In Model B1, the star cluster has almost the same eccentricity as the
previous simulation without SMBH.  In Model B2, the eccentricity of the
star cluster is lower than Model B1.
The orbital elements for the eccentric case were chosen so that the
star cluster would survive at least for several orbits. At first, we
tried the eccentric orbit with the same orbital energy as that of the
circular case. However, in this case, the disruption of the cluster
was too fast, unless the orbit is close to circular. Therefore we made
the eccentric orbit significantly wider.

\onecolumn

\begin{table}[htbp]
\begin{center}
\caption{Models for the galaxy and the star cluster\label{tb:models}}
\begin{tabular}{ccccccccccc}
\tableline
\tableline
& King $W_0$& $N$ & $M({\rm M_{\odot}})$ & $M_{\rm BH}({\rm M_{\odot}})$ & $r_{\rm c}$ (pc) & 
$r_{\rm h}$ (pc) & $r_{\rm t}$ (pc)\\
\tableline
Star cluster & 3 & 65536 & $7.9 \times 10^4$ & - & 0.087 & 0.13 & 0.47 \\
Galaxy 1& 10 & $2 \times 10^6 $ & $8.0 \times 10^7$ & - & 0.66 & 21 & 120 \\
Galaxy 2& 10 & $2 \times 10^6 $ & $2.9 \times 10^7$ & $3.6 \times 10^6$ & - & 9.6 & 72 \\
\tableline
\end{tabular}
\end{center}
\end{table}

\begin{table}[htbp]
\begin{center}
\caption{Initial Conditions for the cluster orbit\label{tb:initial}}
\begin{tabular}{ccccc}
\tableline
\tableline
Simulation & Galaxy model & Orbit & Initial position (pc) & Initial velocity (km/s)\\ \tableline
Model C & 1 & Circular & 2 & 130\\ 
Model E & 1 & Eccentric & 5 & 72 \\
Model B1 & 2 & Eccentric & 5 & 57\\ 
Model B2 & 2 & Eccentric & 5 & 67 \\
\tableline
\end{tabular}
\end{center}
\end{table}

\twocolumn

\subsection{Fully Self-consistent $N$-body Simulation}

We performed fully self-consistent $N$-body simulations using the
Bridge code \citep{Fj07}.  It is a direct-tree hybrid
code. Only the internal motion of the star cluster is calculated by
the direct scheme with high accuracy, and all other interactions are
calculated by the tree algorithm. The splitting between the direct
part and tree part is through the splitting of the Hamiltonian in the
way similar to MVS \citep{WH91,KYN91} or RESPA \citep{Tu90}.
Thus, the low-accuracy calculation of galaxy particles and its
interaction with cluster particles are integrated with time-symmetric
leapfrog algorithm, resulting in small long-term error.  Hence, we can
treat a large-$N$ system with embedded small-scale systems fully
self-consistently and accurately.

The numerical parameters used for the time integration are summarized
in table \ref{tb:param}.  For the tree part, the Bridge code needs the
same parameters with the Barnes-Hut treecode modified for the use with
GRAPE hardware \citep{BH86,M04}.  We used the opening angle $\theta =
0.75$ with the center-of-mass (dipole-accurate) approximation. The
maximum group size for a GRAPE calculation \citep{M91} is 8192.  The
stepsize of leapfrog integrator is $\Delta t = 1/512$ (in Heggie
unit) $= 2.9 \times 10^{-4}$ (Myr) for galaxy 1 and $\Delta t = 1/1024$ 
(in Heggie unit) $= 1.2 \times 10^{-4}$ (Myr) for galaxy 2. The potential is 
softened using Plummer softening.  The softening length between galaxy 
particles that between cluster particles and galaxy particles are the 
same. Both are $\epsilon _{\rm G}=3.9\times 10^{-2}$ pc.
For galaxy 2, we adopted the softening length between between the SMBH
and galaxy particles, $\epsilon _{\rm G-BH}$, is 0.12 pc and between the
SMBH and star cluster particles, $\epsilon _{\rm SC-BH}$, is 0.012 pc.

For the direct part, we used the fourth-order Hermite integrator with
block timestep, and the timestep criterion is of the standard Aarseth
type \citep{MA92} with $\eta = 0.01$. We also used the Plummer softening
for the gravitational force between cluster particles, and we did not
model physical collisions or binary formation in the calculation
reported in this paper. The softening length between star cluster
particles, $\epsilon _{\rm SC}$, is $1.0\times 10^{-5}$ pc.

We stopped the simulations at $T=0.75-0.8$ (Myr). Since
we used the softening and did not model the physical collision and
merging of the stars, the structure of the star cluster after the core
collapse might not expressed correctly. We ignored the stellar
evolution because we treat only very short time ($<$ 1 Myr) in our
simulations.

We used GRAPE-6 \citep{M03} for force calculation.
The total energy was conserved better than $5\times 10^{-5}$ for
the circular orbit $8\times 10^{-5}$ for the eccentric orbit throughout
the simulations.

\begin{table}[htbp]
\begin{center}
\caption{Parameters for $N$-body Simulation\label{tb:param}}  
\begin{tabular}{lcc}
\tableline \tableline
Parameters & Value \\ \tableline
$\epsilon _{\rm G}$ &  $3.9 \times 10^{-2}$ (pc)\\
$\epsilon _{\rm SC}$ & $1.0 \times 10^{-5}$ (pc)\\ 
$\epsilon _{\rm G-BH}$ &  $1.2 \times 10^{-1}$ (pc)\\
$\epsilon _{\rm SC-BH}$ & $1.2 \times 10^{-2}$ (pc)\\
\tableline
$\Delta t$ (for galaxy 1)& $2.9 \times 10^{-4}$ (Myr)\\
$\Delta t$ (for galaxy 2)& $1.2 \times 10^{-4}$ (Myr)\\ \tableline
$\theta$ & 0.75\\
$n_{\rm crit}$ & 8192\\ \tableline
\end{tabular}
\end{center}
\end{table}

\subsection{$N$-body Simulation with Artificial Dynamical Friction}

To compare our result with those of previous works, for model C and E, 
we also performed $N$-body
simulations in which the Galaxy is modeled as a fixed potential and
the dynamical friction due to the Galaxy is calculated analytically.
This treatment is the same as in PZ03, where the acceleration due to
the dynamical friction is calculated using Chandrasekhar's dynamical
friction formula \citep{Ch43,BT87,MP03}
\begin{eqnarray}
\bm{a}_{\rm df} = -4\pi \ln \Lambda \chi G^2 \rho _{\rm G} M_{\rm SC}
 \frac{\bm{v}_{\rm SC}}{v_{\rm SC}^3}\label{eq:df}.
\end{eqnarray}
Here $M_{\rm SC}$ and $\bm{v}_{\rm SC}$ are the mass and the
center-of-mass velocity of the star cluster, $\rho _{\rm G}$ is the
local density of the Galaxy, $\ln \Lambda$ is the Coulomb logarithm,
and 
\begin{eqnarray}
\chi \equiv {\rm erf} (X) - \frac{2X}{\sqrt{\mathstrut \pi}} \exp(-X^2),
\end{eqnarray}
where $X={v_{\rm SC}}/{\sqrt{2} \sigma _{\rm G}}$ and $\sigma _{\rm
G}$ is the local one-dimensional velocity dispersion of the Galaxy,
assumed to be isotropic and locally Maxwellian.  For $M_{\rm SC}$, we
adopted the bound mass, which is the total mass of the particles bound
to the star cluster.  We gave all bound stars the acceleration due
to the dynamical friction from the above formula. Unbound stars were
not affected by the dynamical friction.

In equation (\ref{eq:df}), the Coulomb logarithm, $\ln \Lambda$, is given by
\begin{eqnarray}
\ln \Lambda = \ln \left(\frac{b_{\rm max}}{b_{\rm min}}\right),
\end{eqnarray}
where $b_{\rm max}$ and $b_{\rm min}$ are the maximum and minimum impact
parameters.
It is often set as a constant given by
\begin{eqnarray}
\ln \Lambda \sim \ln \left(\frac{R_{\rm SC}}{\langle r_{\rm SC}\rangle}\right),
\end{eqnarray}
where $R_{\rm SC}$ is the distance of the star cluster from the GC, 
$\langle r_{\rm SC}\rangle$ is the characteristic radius of the star
cluster (roughly the half-mass radius). 
PZ03 adopted a constant value, $\chi \ln \Lambda = 1$.
However, \citet{H03} found a constant $\Lambda$
overestimates dynamical friction at pericenter and proposed a variable
$\Lambda$:
\begin{eqnarray}
\ln \Lambda = \ln \left(\frac{R_{\rm SC}}{1.4 \epsilon_{\rm SC}}\right),
\end{eqnarray}
where $\epsilon _{\rm SC}$ is the size of the star cluster.
We performed $N$-body simulations with artificial dynamical
friction in two ways. One is the
same as that used PZ03 (constant $\Lambda$), and the other is the way
proposed by \citet{H03} (variable $\Lambda$).
We adopted the virial radius of the star cluster for $\epsilon _{\rm SC}$.

\section{Simulation Results}

\subsection{Circular Orbits}

\begin{figure}
\epsscale{0.9}
\plotone{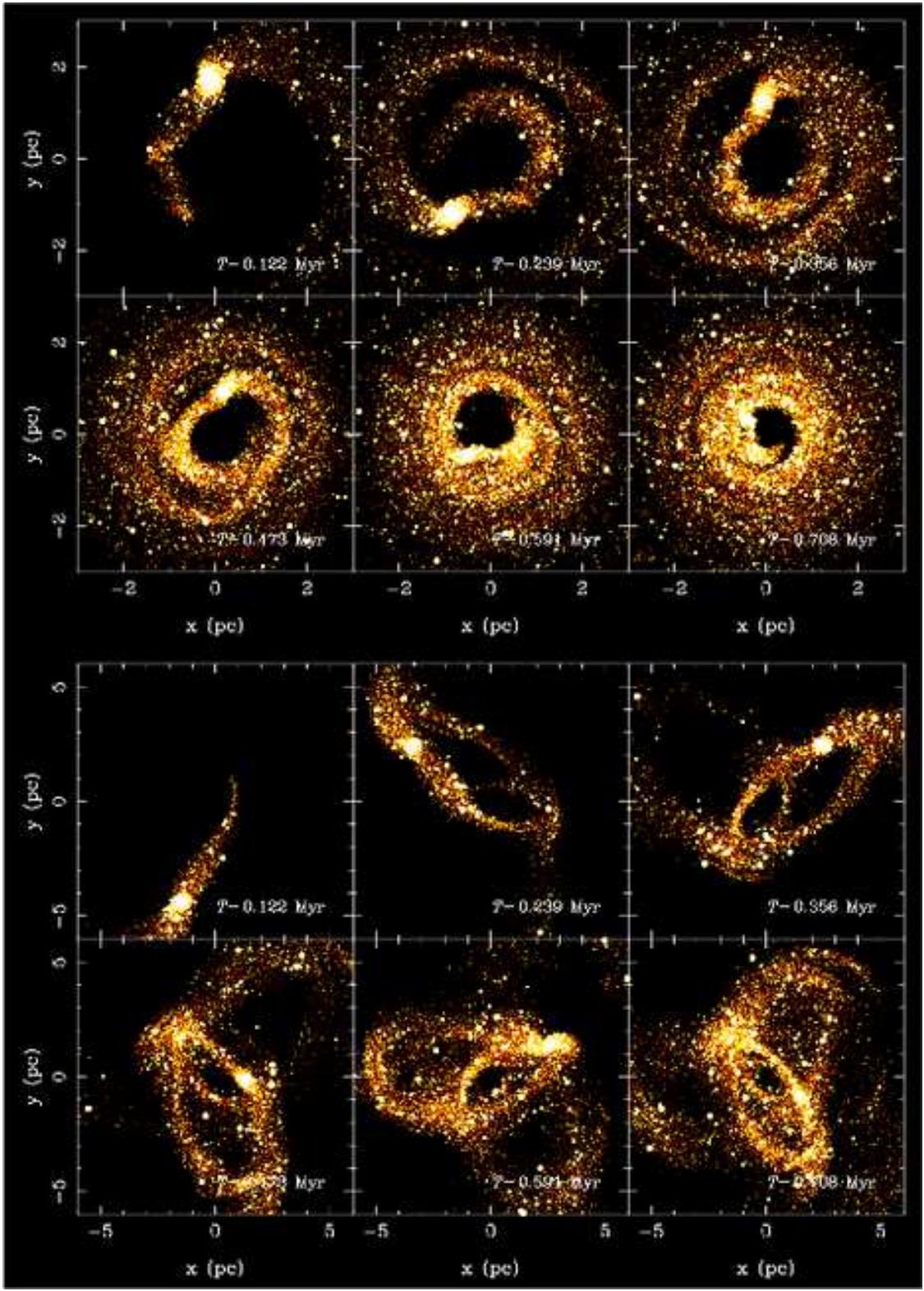}
\caption{Snapshots of the star clusters projected onto $x-y$
plane. The upper six panels are for the run with the circular initial
orbit (model C), and the lower six panels are for the run with the eccentric
orbit (model E). Times are 0.122, 0.239, 0.356, 0.473, 0.591, and 0.708 Myrs.}
\label{fig:snap}
\end{figure}

The top panel of figure \ref{fig:snap} shows the snapshots of the star
clusters projected onto the x-y plane.  The top and bottom panels 
show model C (circular orbit) and E (eccentric orbit), respectively. 
In model C, the star cluster is initially located at 2 pc from the GC. 
Due to the tidal field of the Galaxy, the star cluster
becomes elongated. Particles stripped from the star cluster form tidal
arms and make ring-like structures. Finally they form a disk-like structure.

\begin{figure}
\epsscale{.90}
\plotone{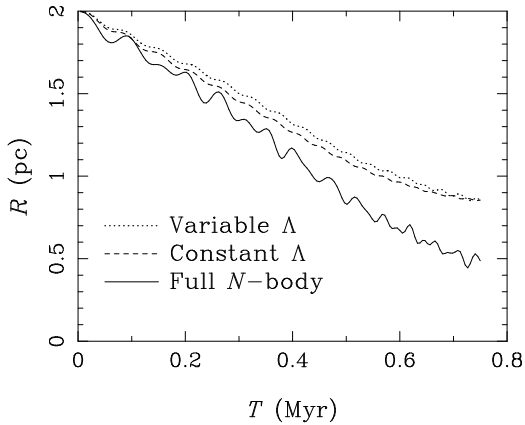}
\caption{The distance of the star cluster from the GC plotted as a
 function of time for model C. 
 Solid curve shows the result of the full $N$-body
 simulation. Dashed and dotted curves show the results of the
 ``traditional'' simulations with variable $\Lambda$ and constant $\Lambda$,
 respectively. 
 \label{fig:radius_cir}}
\end{figure}

\begin{figure}
\epsscale{.90}
\plotone{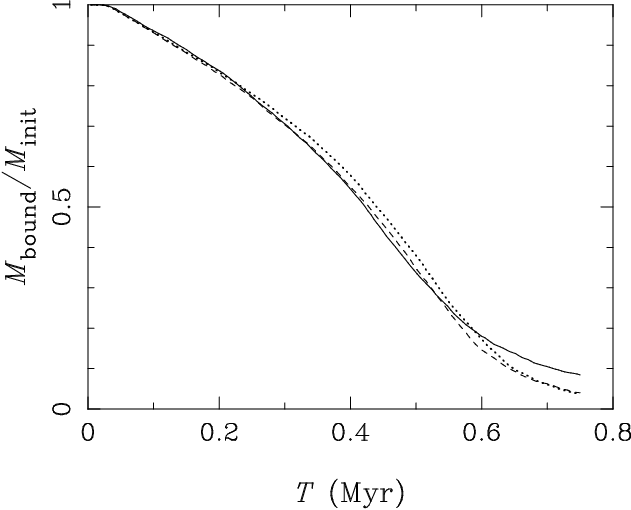}
\caption{The bound mass of the star cluster plotted as a function of
 time for model C. Curves have the same meanings as in
 figure \ref{fig:radius_cir}.\label{fig:mass_cir}}
\end{figure}

We investigated the orbital and internal evolutions of the star
cluster.  Figure \ref{fig:radius_cir} shows the distance of the star
cluster from the GC obtained by our $N$-body simulations.  The solid
curve shows the result of the full $N$-body simulation. The dashed and
dotted curves show the result of the ``traditional'' simulations in
which the dynamical friction is calculated analytically from the
formula with constant $\Lambda$ and variable $\Lambda$, respectively.
The orbital decay in the full $N$-body simulation is faster by 30-40\%
than that in other simulations. On the other hand, the evolution of
the bound mass of the star cluster shown in figure \ref{fig:mass_cir}
is almost the same among the three.  This result suggests that previous
studies underestimated the inspiral timescale of star clusters. This
effect is not very large, but not negligible.

In figure \ref{fig:mass_cir}, it seems that the mass loss in the very
late stage (after more than 80\% of the initial total mass is lost)
seems to be significantly slower for the case of full $N$-body
simulation than those for other two simulations, even though the cluster
is much closer to GC. This difference is probably due to the finite core
size of our galaxy model ($r_c = 0.66{\rm pc}$) and not the reality. In
order to study the evolution after this stage, we need to use a more
realistic model of the mass distribution of the central parsec of the
galaxy, which includes the central massive black hole. In addition,
stellar collision and merger within the star cluster must be modeled.
We are currently working on such an extension of the Bridge code.

\begin{figure}
\epsscale{.90}
\plotone{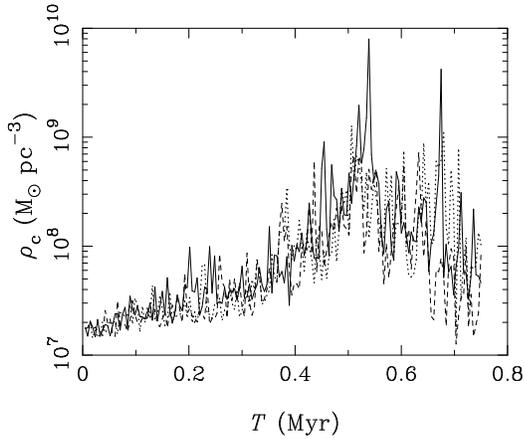}
\caption{The core density of the star cluster, $\rho _{\rm c}$,
 plotted as a function of time for model C. Curves have the
 same meanings as in figure \ref{fig:radius_cir}.\label{fig:density_cir}}
\end{figure}

Figures \ref{fig:core_cir} and \ref{fig:density_cir} show the core
radius and the core density of the star cluster.  The core collapse
occurred at around $T\simeq 0.53$ Myr.  The core collapse times are
also the same among the three simulations.

\onecolumn
\twocolumn

\begin{figure}
\epsscale{.90}
\plotone{f5.eps}
\caption{The core radius of the star cluster, $r_{\rm c}$,  plotted as a
 function of time for model C. Curves have the same
 meanings as in figure \ref{fig:radius_cir}.\label{fig:core_cir}}
\end{figure}

\subsection{Eccentric Orbits}

The bottom panel of figure \ref{fig:snap} shows the snapshots of 
model E projected onto the x-y plane. The initial distance of
the star cluster from the GC is 5 pc.  The star cluster is elongated
and particles are stripped due to the tidal force of the Galaxy. The
stripped particles form complex tidal tails.

\begin{figure}
\epsscale{.90}
\plotone{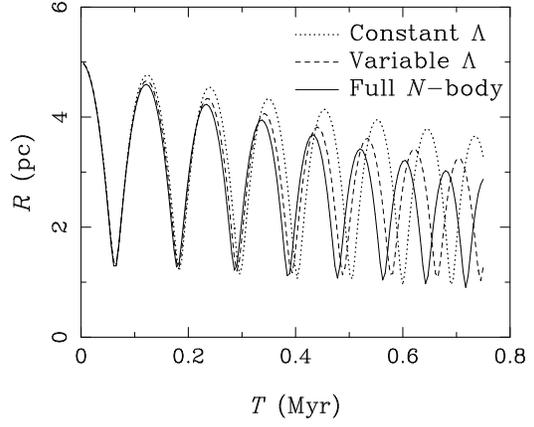}
\caption{The distance of the star cluster from the GC plotted as a
 function of time for model E.   
 Solid curve shows the result of the full $N$-body
 simulation. Curves have the same meanings as in
 figure \ref{fig:radius_cir}.\label{fig:radius_ecc}}
\end{figure}

\begin{figure}
\epsscale{.90}
\plotone{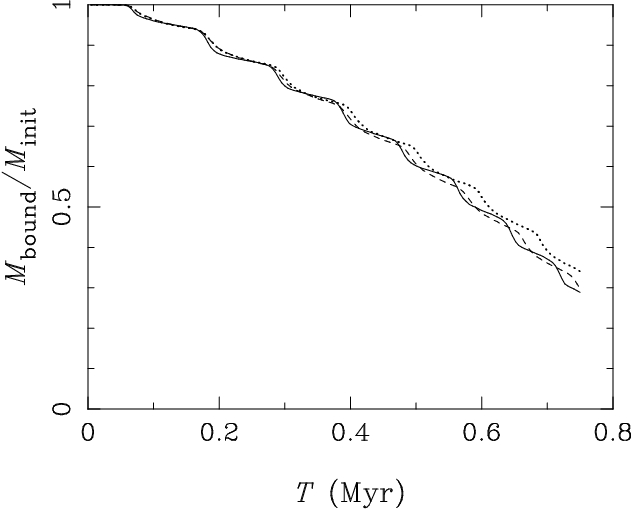}
\caption{The bound mass of the star cluster plotted as a function of
 time for model E. Curves have the same meanings as in
 figure \ref{fig:radius_ecc}.\label{fig:mass_ecc}}
\end{figure}

Figure \ref{fig:radius_ecc} shows the orbital evolution of the star
cluster.  The orbital decay of
the full $N$-body simulation is faster than the traditional
simulations, as was the case in the runs from the circular orbit. In
this case also, the evolution of the bound mass is the same among the
three simulations (see figure \ref{fig:mass_ecc}).  
Moreover, this
result shows that variable $\Lambda$ works better than constant
$\Lambda$.

\begin{figure}
\epsscale{.90}
\plotone{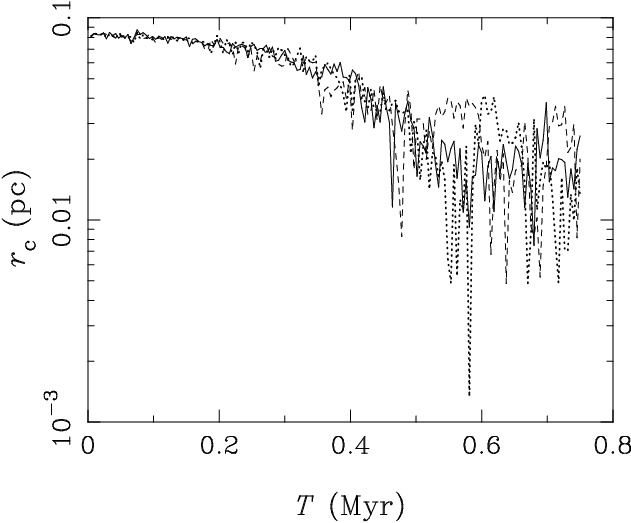}
\caption{The core radius of the star cluster, $r_{\rm c}$, plotted as a
 function of time for model E. Curves have the same
 meanings as in figure \ref{fig:radius_ecc}.\label{fig:core_ecc}}
\end{figure}

\begin{figure}
\epsscale{.90}
\plotone{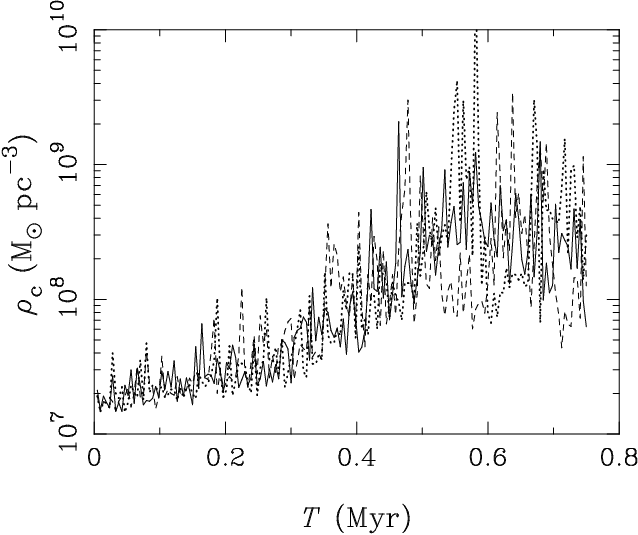}
\caption{The core density of the star cluster, $\rho _{\rm c}$,
 plotted as a function of time for model E. Curves have the same
 meanings as in figure
 \ref{fig:radius_ecc}.\label{fig:density_ecc}}
\end{figure}

Figures \ref{fig:core_ecc} and \ref{fig:density_ecc} show the core radius
and the core density of the star cluster.
These show that the core collapse occurred at around $T\simeq 0.55$
Myr. The core collapse time is almost the same as that for the case of
the circular orbit, even though the bound mass at the collapse time is
different by almost a factor of two. The half-mass relaxation time at
the time of the collapse was 0.11 Myr and 0.34 Myr, in 
model C and E, respectively. Thus, if clusters are rapidly
losing its mass due to the tidal field, the apparent age measured by
the present relaxation time can show large variations, even if they
started from the same initial condition and collapsed at the same time.
Furthermore, the core collapse times are the same as that in the
case without massloss due to the tidal field. The core collapse 
time, $t_{\rm cc}$, is estimated as $t_{\rm cc} \simeq 0.20 t_{\rm rh}$,
where $t_{\rm rh}$ is half-mass relaxation time \citep{PM02}. From this
equation, we obtain the core collapse time of our model star cluster as
0.51 Myr.  Our results agreed with it.

\begin{figure}
\epsscale{.90}
\plotone{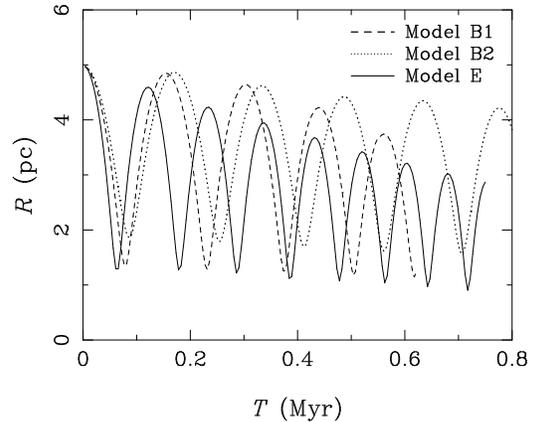}
\caption{The distance of the star cluster from the GC plotted as a
 function of time. Solid curve shows the result of the galaxy model
 without the central SMBH (Model E; the same as the solid curve in figure
 \ref{fig:radius_ecc}). Dashed and dotted curves show the result of the
 galaxy model with the central SMBH (Model B1 and B2).\label{fig:radius_BH}}
\end{figure}

\begin{figure}
\epsscale{.90}
\plotone{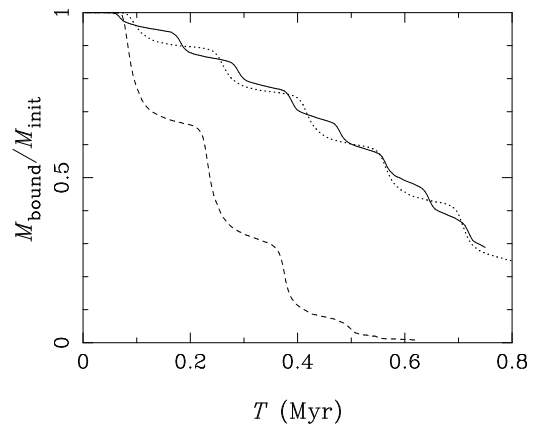}
\caption{The bound mass of the star cluster plotted as a function of
 time for the galaxy model with the SMBH. Curves have the same meanings as in
 figure \ref{fig:radius_BH}.\label{fig:mass_BH}}
\end{figure}

\begin{figure}
\epsscale{.90}
\plotone{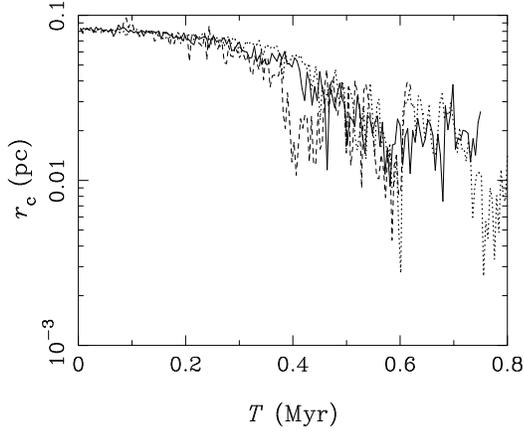}
\caption{The core radius of the star cluster, $r_{\rm c}$,  plotted as a
 function of time for the galaxy model with the SMBH.  Curves have the same
 meanings as in figure \ref{fig:radius_BH}.\label{fig:core_BH}}
\end{figure}

\begin{figure}
\epsscale{.90}
\plotone{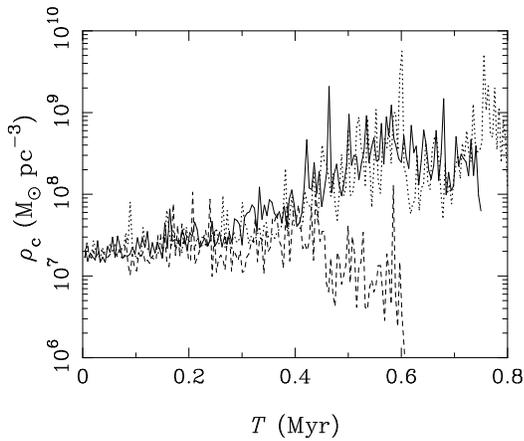}
\caption{The core density of the star cluster, $\rho _{\rm c}$, plotted
 as a function of time for the galaxy model with the SMBH. Curves have the
 same meanings as in figure \ref{fig:radius_BH}.\label{fig:density_BH}}
\end{figure}

Figure \ref{fig:radius_BH} - \ref{fig:density_BH} shows the results of 
model B1 (dashed curves) and B2 (dotted curves), where the Galaxy model 
has a central BH.  Solid curves show the result of model E (no BH; the same 
as the solid curve in figure \ref{fig:radius_ecc} - \ref{fig:density_ecc}).  
The orbital evolution of the star cluster in model B1 and B2 is essentially 
similar to that in model E.  However, the orbital decay in model
B1 is somewhat slower than that in model E.  As is clear
in figure \ref{fig:mass_BH}, in model B1, the mass loss at the
pericenter passage is much larger than in the case of model E.
Therefore, the core collapse did not occur and the core density of the 
cluster did not increase.  
This is because the star cluster suffers the strong tidal
force from the SMBH at the pericenter.  The small mass of the star
cluster slows the orbital evolution of the star cluster.  Furthermore,
the strong tidal field disrupts the star cluster much faster than the
case without SMBH.  

In model B2, the pericenter of the star cluster is farther than that in
model B1.  The evolution of the bound mass is the same as the case
without SMBH.  However, the orbital evolution is much slower than that
of model B1, because the pericenter of the star cluster is farther.

The cluster model and its orbital evolutions in model E and B1 are 
almost the same as simulation 8 of KM03.
Their model of the Galaxy has a power-law density profile with the
central SMBH, while our galaxy model 1 for model E has no SMBH.  However, 
the enclosed mass of the galaxy in KM03 is very similar to ours between
around 1 and 10pc.  In their simulation, the star cluster was totally
disrupted before 1 Myr, but in our simulation, the star cluster has
30\% of its initial mass at the end of our simulation (0.75 Myr).  The
difference is caused by the internal evolution of the star cluster.
In our simulation, mass segregation and core collapse of the cluster
made the central density much higher, which prevent the complete
disruption of the cluster.  On the other hand, in KM03, such an
evolution was prevented by the numerical method they used.

On the other hand, the cluster in model B1 was disrupted before the 
core collapse occurs by the strong tidal field of the central SMBH.  
At around 1 pc, the enclosed mass of galaxy model 2 which includes a SMBH 
is twice as large as that of K04's model.  Such a difference of the 
enclosed masses is caused by the mass of the SMBH.  We adopted 
$3.6\times10^6 M_{\sun}$ \citep{E05}, while KM03 did $2.5\times10^6 M_{\sun}$.

Thus, the presence of a SMBH has considerable effect on the evolution of
star clusters.  The main effect is that the tidal field near the GC
becomes much stronger, resulting in faster mass loss from star
clusters and preventing its core collapse.  
In the models we tested, the core collapse time is about the same as the 
time of the complete disruption in model B1.  If the core collapse time is
somewhat faster, it is possible that the cluster survives and approaches
to the GC. In particular, if an IMBH is formed within the star cluster, it
would significantly help the survival of the cluster.  We will study this
aspect in more details in the forthcoming papers.

\subsection{Eccentricities and Inclinations of the Escaped Stars}

For some of the stars in the central parsec, the projected positions
and the proper motions have been measured, and their orbital elements
have been estimated \citep{Pa06,Lu06}. Young and bright stars
apparently belong to one of the two ``disks'' (clockwise and
counter-clockwise rotating disks), though the existence of the
counter-clockwise disk is controversial \citep{Lu06}.  The
eccentricities of the stars on the counter-clockwise rotating disk are
high, $\sim 0.8$ \citep{Pa06}. For the stars on the clockwise rotating
disk, \citet{Pa06} concluded their orbits are circular, while
\citet{Lu06} concluded the lower limits of their eccentricities
distribute between 0.0 and 0.8. These high eccentricities are
difficult to explain with the in-situ formation scenario, and also have
been thought to be difficult to explain with cluster inspiral
scenario, since in both cases the stars would have close-to-circular
orbits. 

\begin{figure}
\epsscale{1.0}
\plotone{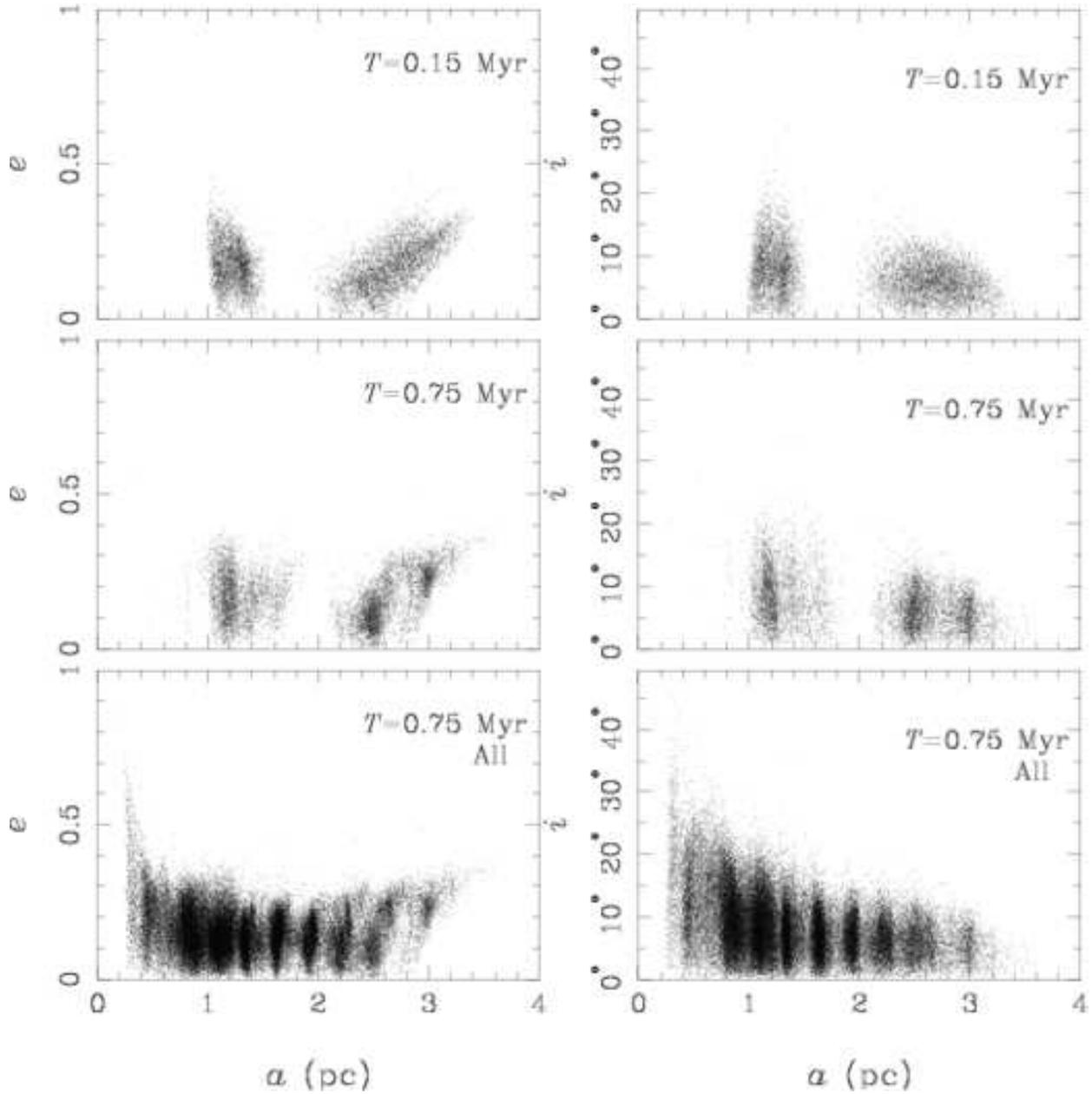}
\caption{The eccentricity (left) and inclination (right) of the stars
 escaped from the star cluster as the function of their the
semi-major axis $a$ for model C (circular orbit).
 The top panels show the unbound stars at $T=0.15$ Myr and the
 middle panels show the position of the same stars shown in top
 panels, but at $T=0.75$ Myr. The bottom panels show the all unbound
 stars at $T=0.75$ Myr.
 \label{fig:ei_cir}}
\end{figure}

\begin{figure}
\epsscale{1.0}
\plotone{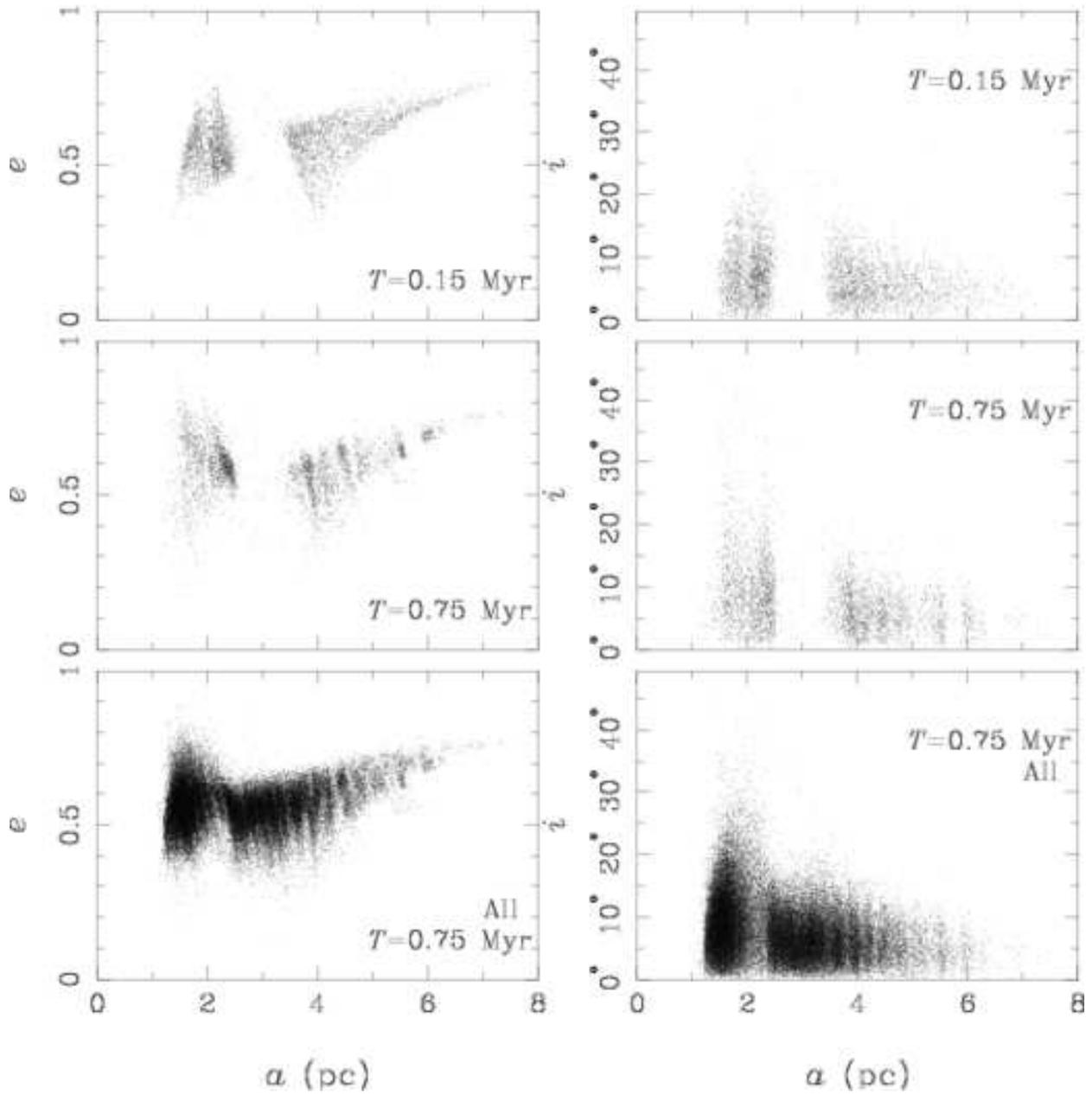}
\caption{Same as figure \ref{fig:ei_cir}, but for model E (eccentric
 orbit). \label{fig:ei_ecc}}
\end{figure}

We investigated the eccentricities, $e$, and inclinations, $i$, of
stars escaped from the star cluster (i.e. unbound stars) for model C and
E.  Figures \ref{fig:ei_cir} and \ref{fig:ei_ecc} show the
eccentricities and inclinations of escaped stars and their evolutions
for model C and E, respectively.  In these figures, the $x$-axis is the
semi-major axis, $a$, of the star.  The eccentricity is defined as 
$e = (r_{\rm a}-r_{\rm p})/(r_{\rm a}+r_{\rm p})$, where $r_{\rm a}$ and
$r_{\rm P}$ are the apocenter and pericenter distances, respectively. We
obtained the peri- and apocenter distances by integrating the orbits of
the stars
in the model potential. The inclination is defined as 
$i=\cos ^{-1} (h_{z}/h)$, where $h$ and $h_{z}$ are the angular momentum
and its $z$ component, respectively.  The top panels show the
eccentricities and inclinations of the stars escaped before $T=0.15$
Myr. The central gap corresponds to the semi-major axis of the star
cluster, and the left and right wings are the stars on
the leading and trailing arms, respectively. The distributions of the
escaped stars on the $a-e$ and $a-i$ plains expand because of the
time-varying gravitational force from 
the star cluster (see the middle panels). The bottom panels show the
distribution of all escaped stars at the last snap shots, $T=0.75$
Myr. Even in the case of the circular orbit (model C), the
eccentricities of the escapers in innermost orbits can reach rather
large values. However, in the case of the eccentric orbit (model E), the
eccentricities of the escapers are even higher. They distribute between
0.4 and 0.8, being consistent with the observed values of stars in the
two ``disks''.  Thus, if the cluster is initially in a highly eccentric
orbit, the high eccentricities of observed stars are naturally explained.

\onecolumn

\twocolumn

\subsection{The Evolution of the Mass Function}

\begin{figure}
\epsscale{.90}
\plotone{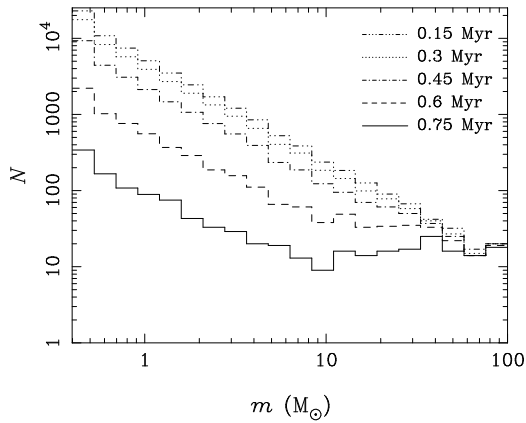}
\caption{The mass function of the bound stars for model C. Times are
 $T=$ 0.15, 0.3, 0.45, 0.6, and 0.75 Myrs.\label{fig:MF1}}
\end{figure}

\begin{figure}
\epsscale{.90}
\plotone{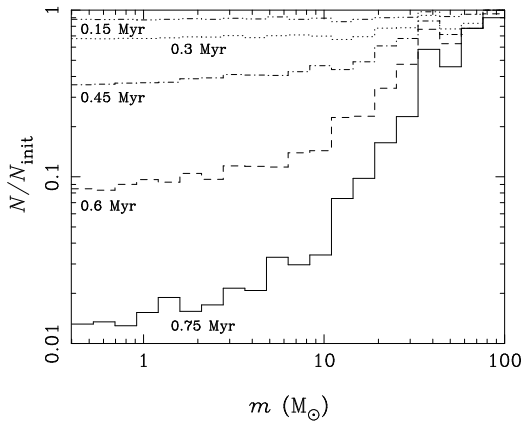}
\caption{The fraction of the mass function of the bound stars to the
 initial mass function for model C. Times are the same as figure
 \ref{fig:MF1}.\label{fig:MF}}
\end{figure}

Figure \ref{fig:MF1} shows the evolution of the mass function (MF) of
the stars bound to the star cluster for model C. 
We can see that the slope of the mass function for $m> 10M_{\odot}$
becomes flatter as the system evolves, while that for $m< 10M_{\odot}$
shows rather little change. To see this effect more clearly, in figure
\ref{fig:MF} we plot the fraction of the mass retained in the cluster
as the function of the stellar mass for model C.  
Stars with mass $m> 30M_{\odot}$
are almost perfectly retained in the cluster, and the retention rate
quickly drops in the range of $10M_{\odot}<m< 30M_{\odot}$. For 
 $m< 10M_{\odot}$, retention rate becomes smaller for smaller mass,
but the dependence becomes much weaker.

\begin{figure}
\epsscale{.90}
\plotone{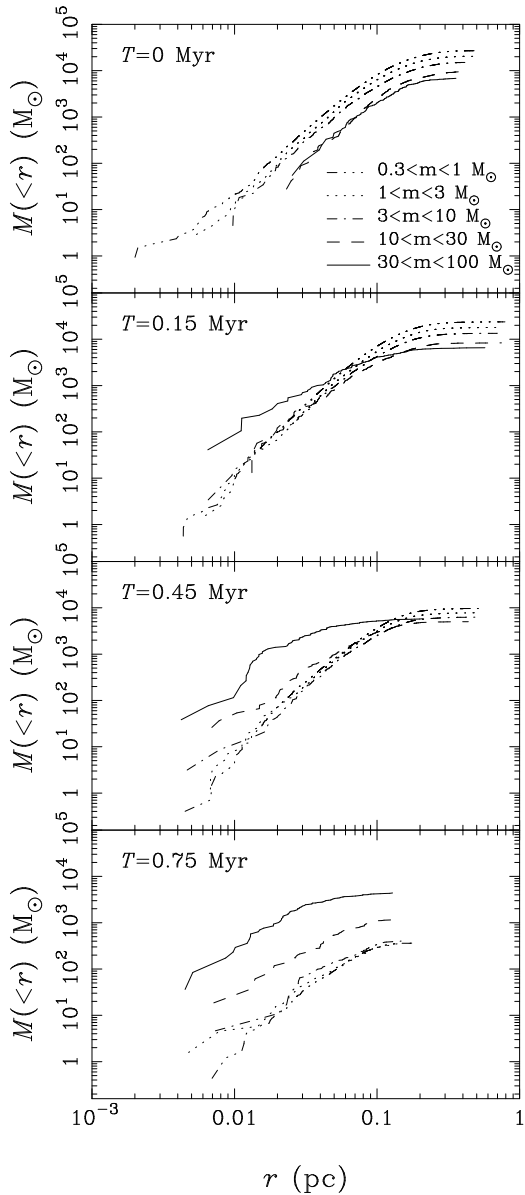}
\caption{The evolution of the enclosed mass of the bound stars for model
 C.\label{fig:rm}}
\end{figure}

Figure \ref{fig:rm} shows the evolution of the enclosed mass in $r$,
the distance from the center of the star cluster for model C. 
The enclosed masses are calculated for five mass ranges. 
Initially, all mass ranges have same profile. At $T=0.15$ Myr, the most
massive stars have sank to the center and the second massive stars also
shows some central condensation.  At $T=0.45$ Myr, a large fraction of
stars more massive than 10 $M_{\sun}$ have sank to the center (within
radius 0.02 pc). However, the distribution of stars with mass less than 
$10 M_{\sun}$ have not changed significantly. Thus, the stars heavier
than 10 $M_{\sun}$ remained in the star cluster.

We can estimate the critical mass of the star at which this change in
the behavior occurs, by calculating the mass
of the star at which the dynamical friction just balances the two-body
heating.  First, we consider the energy change of the stars in the
system consisting of two components with the masses $m_1$ and $m_2$.
Assuming that the velocity dispersions of both components are
Maxwellian, the mean energy change of a star of mass $m_1$ is
expressed  as
\begin{eqnarray}
\left< \frac{d}{dt}(m_1E_1) \right> = \frac{4\sqrt{3\pi}G^2m_1m_2n_2\log
 \Lambda}{(\left< E_1\right> +\left< E_2\right>)^{2/3}}
(m_2\left< E_2\right> - m_1\left< E_1 \right>),\label{eq:energy1}
\end{eqnarray}
where $\left< E_1\right>$ and $\left< E_2\right>$ are mean specific
kinetic energy of each components, $n_2$ is the number density of the
stars of mass $m_2$, and $\ln \Lambda$ is the Coulomb logarithm
\citep{HH03}.  At least in the initial model, the velocity dispersion
of the stars is independent of the mass. So we can set 
$\left< E_1\right> = \left< E_2\right> = 1/2 \sigma ^2$. 
In this case, equation (\ref{eq:energy1}) is rewritten as
\begin{eqnarray}
\left< \frac{d}{dt}(m_1E_1) \right> = An_2m_1m_2(m_2-m_1),\label{eq:energy2}
\end{eqnarray}
where $A$ is a constant.  Now we consider the case of continuous mass
distribution, in which the mass distribution of $m_2$ is given by
\begin{eqnarray}
\frac{dn}{dm} \propto m^{-\alpha},
\end{eqnarray}
where $C$ is a constant.
The number density of mass $m_2$, $n_2$, is expressed as
 \begin{eqnarray}
n_2 = C m_2^{-\alpha}.\label{eq:MF}
\end{eqnarray}
By substituting equation (\ref{eq:MF}) to (\ref{eq:energy2}) and
integrating it over mass $m_2$, we can obtain the energy change of
stars with mass $m_1$ as 
\begin{eqnarray}
\left< \frac{d}{dt}(m_1E_1) \right> &=& A'm_1\int^{m_{\rm max}}_{m_{\rm min}}m_2^{-\alpha+1}(m_2-m_1)dm_2\\
&\propto&m_1 \left[\frac{m_2^{-\alpha +3}}{-\alpha +3} -\frac{m_2^{-\alpha+2}}{-\alpha+2}m_1\right]_{m_{\rm min}}^{m_{\rm max}},\label{eq:energy3}
\end{eqnarray}
where $m_{\rm max}$ and $m_{\rm min}$ are the maximum and minimum mass of
the MF and $A'$ is a constant.
If the right side of equation (\ref{eq:energy3}) is negative, the star
with mass $m_1$ loses energy and sink to the center of the star cluster.
The minimum mass with the negative energy change, $m_{\rm sink}$, is
expressed as 
\begin{eqnarray}
m_{\rm sink}&=&\frac{-\alpha +2}{-\alpha +3} \hspace{3pt}
\frac{m_{\rm max}^{-\alpha+3}-m_{\rm min}^{-\alpha+3}}{m_{\rm
max}^{-\alpha+2}-m_{\rm min}^{-\alpha+2}}\\
&=&f(\alpha) m_{\rm max}\frac{1-x^{-\alpha +3}}{1-x^{-\alpha +2}},
\end{eqnarray}
where we defined $x\equiv m_{\rm min}/m_{\rm max}$ and
$f(\alpha) \equiv (-\alpha +2)/(-\alpha +3)$.
Using the values used for our model, $m_{\rm max}=100M_{\sun}$, 
$m_{\rm min}=0.3M_{\sun}$ and $\alpha = 2.35$, we obtain 
$m_{\rm sink}=7.9M_{\sun}$.
This value agrees well with the mass where the power-low index of the MF
breaks in figure \ref{fig:MF}.
If $x \ll 1$, we obtain
\begin{eqnarray}
\frac{1-x^{-\alpha +3}}{1-x^{-\alpha +2}}=
\left\{ \begin{array}{ll}
1 & (\alpha > 3),\\
-x^{\alpha-2} & (2<\alpha < 3),\\
x & (\alpha < 2).
\end{array} \right.
\end{eqnarray}
Therefore, 
\begin{eqnarray}
m_{\rm sink} &\simeq&
\left\{ \begin{array}{ll}
f(\alpha) m_{\rm min} & (\alpha > 3),\\
-f(\alpha) m_{\rm min}^{\alpha -2} m_{\rm max}^{3-\alpha} & (2<\alpha < 3),\\
f(\alpha) m_{\rm max} & (\alpha < 2).
\end{array} \right.
\end{eqnarray}
When $2<\alpha <3$, the value of $m_{\rm sink}$ varies from 
$m_{\rm min}$ to $m_{\rm max}$.

The observed MF of the stars in the central parsec is much flatter
than Salpeter \citep{Pa06}. The cluster inspiral model rather
naturally explain this flat MF, since only the most massive
stars remain bound to the cluster and are carried to the central region of
the galaxy.

\section{Summary and Discussion}

\subsection{Summary}

We performed fully self-consistent $N$-body simulations of a star
cluster within its parent galaxy and compared the orbital and internal
evolutions of the star cluster with those obtained by ``traditional''
simulations, in which the orbital evolution of the star cluster is
calculated from the dynamical friction formula.
We confirmed that the inspiral timescale of the star cluster is shorter
than that obtained from the ``traditional'' simulations.  Furthermore, 
our results showed that the core collapse make the core density of the 
cluster increase and helps the cluster survive.  

We performed simulations of circular and eccentric orbits of the star
cluster.  In previous studies, most of the simulations were from
circular orbits (PZ03; G\"{u}rkan \& Rasio 2005).  We found that,
however, eccentric orbits are more favorable to explain the distribution
of stars around the GC for following reasons.

First, eccentric orbits are natural, if the formation of star clusters
was triggered by collisions between gas clouds.  Second, star clusters
with eccentric orbits can approach to the GC much faster than those with
circular orbits (KM03).  Third, \citet{Pa06} showed that many stars in the
counterclockwise rotating disk have high eccentricities
($e\simeq 0.8$), while the distribution of the eccentricities
of the stars in the clockwise disk is very broad. Since the
eccentricities of the escaped stars distribute around the eccentricity
of the star cluster, the star cluster model with eccentric orbits can
naturally explain the existence of high-eccentricity stars.

The power-low index of the MF of the bound stars to the star cluster
breaks at around 7.9 $M_{\sun}$. Stars heavier than this mass sink to
the center of the star cluster due to the mass segregation. Since the
tidal stripping removes the stars outside of the star cluster, the
massive stars selectively remain in the star cluster. As a result, the
star cluster carries only massive stars to the GC.  The star cluster
scenario can reproduce the flat MF in the central parsec, without the
need for nonstandard initial mass function.

\subsection{Realistic Model of the Galaxy}
First, we showed the orbits of star clusters decays faster in
full $N$-body simulations than in ``traditional'' simulations, which
treat the dynamical friction analytically using a King model $W_{0}=10$
as a model of the central region of the Galaxy. 
The model is sufficient for such comparisons, but not for more realistic
comparison between our model and the stars in the GC because within
$\sim$1 pc the mass density of our model is much lower than that of
the actual Galaxy.  
Next, we showed the case of more realistic galaxy model with a central
BH.  The orbital evolution was similar that in
the model without a BH.  For the comparison between our simulations and
the actual stars, however, we need more simulations in various initial
conditions.  We will report more detail result of runs with the central
SMBH in forthcoming papers.

\subsection{Formation of an Intermediate-mass Black Hole (IMBH)}

IRS 13E consists of seven stars within a projected diameter of $\sim$
0.02 pc and is located at $\sim$ 0.14 pc in projection from the GC
and these stars have very similar proper motions \citep{Maillard04,Pa06}. 
\citet{Maillard04} suggested that IRS 13E is the remnant core of a star
cluster that have fallen to the GC and dissolved there and that the members
of IRS 13E are bound by a central IMBH.
From the analysis of proper motions, the minimum mass of the IMBH was
estimated as 1300 $M_{\sun}$ \citep{Maillard04} - $10^4 M_{\sun}$
\citep{Schodel05}.

In this paper we simulated the evolution of star clusters only before
the core collapse because of the limitation of our present code.  Our
code currently cannot treat the post-collapse evolution since we use the
softened potential. Collisions and mergers between stars would have
occurred and an IMBH would have be formed, if our code can handle these
events. The star cluster inspiral scenario might reveal the origin of
the IRS 13E. We are currently working to implement collisions and
mergers. The result will be reported in the future papers.

\acknowledgments

The authors thanks Piet Hut, Keigo Nitadori, and Ataru Tanikawa for
useful comments and discussions.
M. F. is financially supported by Research Fellowships of the Japan
Society for the Promotion of Science (JSPS) for Young Scientists.
This research is partially supported by 
the Special Coordination Fund for Promoting Science and Technology
(GRAPE-DR project), Ministry of Education, Culture, Sports, Science and
Technology, Japan.
Part of calculations were done using the
GRAPE system at the Center for Computational Astrophysics (CfCA) of
the National Astronomical Observatory of Japan.

\end{document}